\documentclass[twocolumn,showpacs,preprintnumbers,amsmath,amssymb]{revtex4}
\usepackage{graphicx}
\usepackage{dcolumn}
\usepackage{bm}

\begin{document}

\preprint{APS}

\title{Elastic backbone defines a new transition in the percolation model}

\author{Cesar I. N. Sampaio Filho$^1\footnote{Correspondence to: cesar@fisica.ufc.br}$, Jos\'{e} S. Andrade Jr.$^{1,2}$, Hans J. Herrmann$^{1,2}$, Andr\'{e} A. Moreira$^1$}

\affiliation{$^1$Departamento de F\'{i}sica, Universidade Federal do Cear\'a, 
  60451-970 Fortaleza, Cear\'a, Brasil\\
  $^2$Computational Physics for Engineering Materials, IfB, ETH Zurich, 
  Schafmattstrasse 6, 8093 Zurich, Switzerland}

\begin{abstract}
The elastic backbone is the set of all shortest paths. We found a new phase transition at $p_{eb}$ above the classical percolation threshold at which the elastic backbone becomes dense. At this transition in $2d$ its fractal dimension is $1.750\pm 0.003$, and one obtains a novel set of critical exponents $\beta_{eb} = 0.50\pm 0.02$, $\gamma_{eb} = 1.97\pm 0.05$, and $\nu_{eb} = 2.00\pm 0.02$ fulfilling consistent critical scaling laws. Interestingly, however, the hyperscaling relation is violated. Using Binder's cumulant, we determine, with high precision, the critical probabilities $p_{eb}$ for the triangular and tilted square lattice for site and bond percolation. This transition describes a sudden rigidification as a function of density when stretching a damaged tissue.
\end{abstract}

                                          
\maketitle

Despite being one of the simplest and most studied models, classical percolation \cite{broadbent1957percolation,kirkpatrick1973percolation,stauferbook1985,sahimiBook1994} still bears yet uncovered surprises. It is well known that at the percolation threshold ($p_c$) the shortest path between two opposite sides of the system is fractal, with a fractal dimension that is only known numerically to be $d_{sp} = 1.1307\pm 0.0004$ \cite{herrmann1988fractal,grassberger1992,havlinJStat1998,ziffPRL2000} in two dimension and increases with dimension, until  becoming two at and above the critical dimension $d=6$. In fact the shortest path is not unique: several shortest paths can exist simultaneously and the set of all shortest paths has been called “elastic backbone” in the past \cite{herrmann1984backbone}, because it is the subset of the backbone that, when elongated, would give the first contribution to a restoring force. The elastic backbone indeed determines the first resistance that is felt, when stretching damaged~\cite{thorpePRL1995,souzaPNAS2009,PhysRevMaterials} or biological tissues \cite{bates1994lung,yuan1997dynamic,baish2000fractals,suki2002fluctuations,ritter2009zipper, lennon2015lung} and thus has experimental relevance. It has been numerically established that at the percolation threshold $p_c$ its fractal dimension is indistinguishable from the one of the shortest path \cite{herrmann1984backbone}. Here we report on the discovery that, above the classical percolation threshold $p_c$, there exists another critical probability $p_{eb} >  p_c$ at which the elastic backbone becomes
dense. While its dimension is known to be $d_{sp}$ at $p_c$, our results in two dimensions show that it becomes unity between $p_{c}$ and $p_{eb}$, it is $1.750\pm 0.003$ at $p_{eb}$, and is equal to two above. At $p_{eb}$ we reveal critical scaling laws and a set of exponents, however, violating hyperscaling.

\begin{figure}[t]
\includegraphics*[width=5.0cm,height=10.0cm]{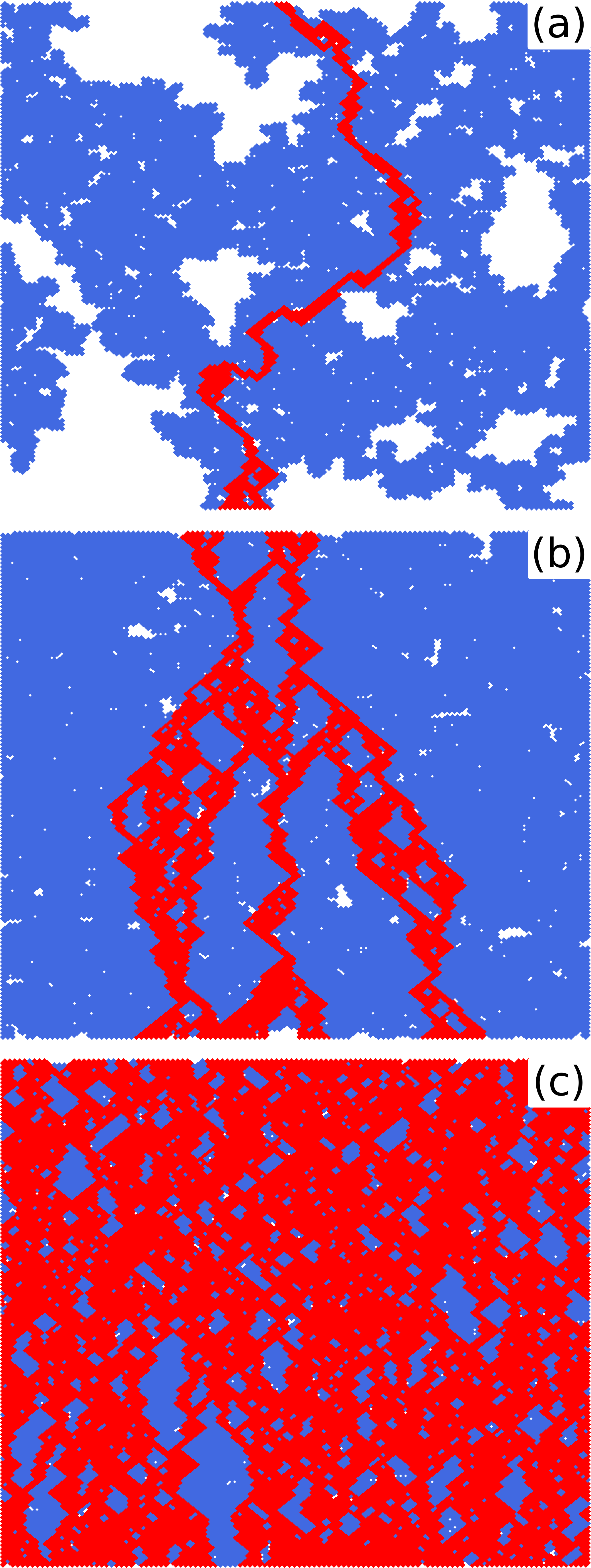}
\caption{Shown in red are typical elastic backbones obtained for site percolation on a tilted square lattice of linear size $L=512$ and calculated with: (a) $p=0.6000$ ($p_{c}<p<p_{eb}$), (b) $p=0.7055$ ($p=p_{eb}$), and (c) $p=0.7500$ ($p>p_{eb}$). Occupied sites that are in the spanning cluster, but do not belong to the elastic backbone, are shown in blue. Other sites are represented in white.}
\label{fig01}
\end{figure}

We simulated two-dimensional percolation configurations at occupation probability $p$ for systems of size $L\times L$ with periodic boundary conditions in horizontal direction and open boundaries at top and bottom. Using the burning algorithm, we identified for each configuration the elastic backbone, i.e. the set of all shortest paths \cite{herrmann1984backbone}. We considered site and bond percolation on the triangular lattice and on two types of square lattices, namely, tilted and non-tilted. In Fig.~\ref{fig01} we see the elastic backbones for site percolation on a tilted square lattice for three different probabilities below, at and above $p_{eb}$.      

\begin{figure}[t]
\includegraphics*[width=\columnwidth]{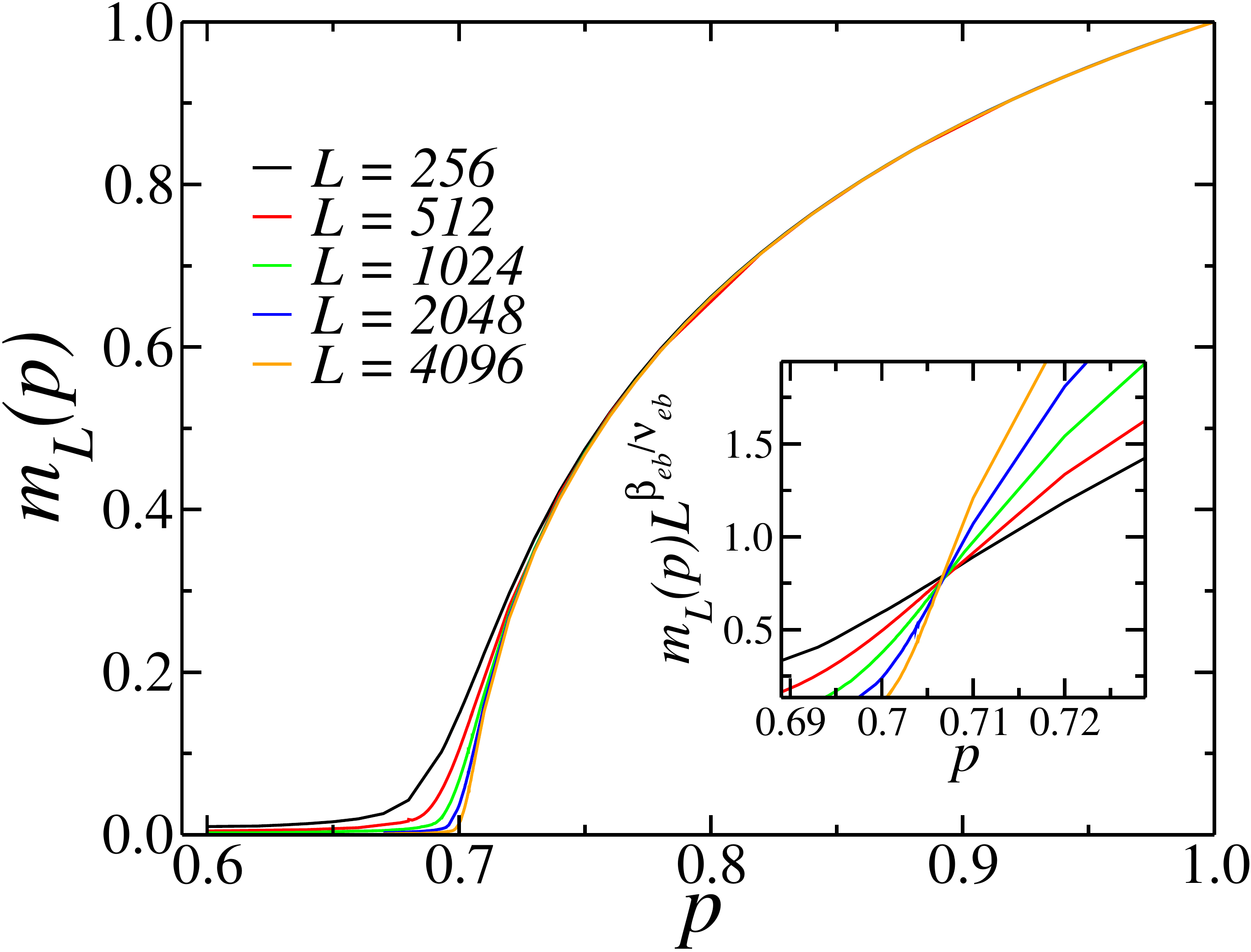}
\caption{Density $m_{L}(p)$ of the elastic backbone as a function of the occupation probability $p$ for site percolation on the triangular lattice for different system sizes. Inset: determination of the threshold using finite-size scaling.}
\label{fig02}
\end{figure}

\begin{figure}[t]
\includegraphics*[width=\columnwidth]{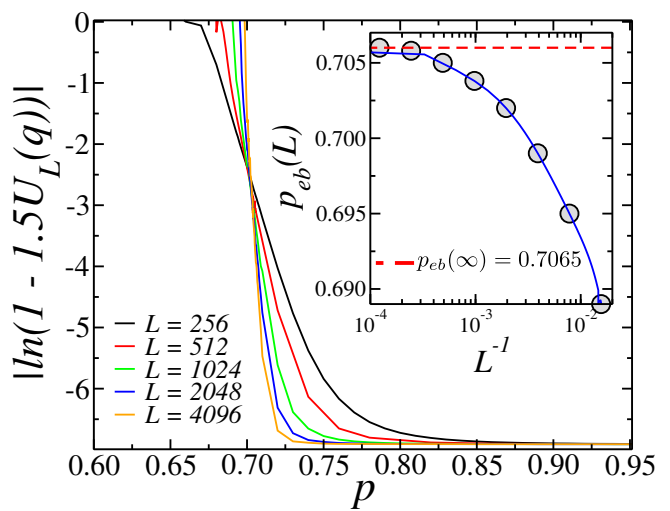}
\caption{Binder's cumulant $U_{L}(p)$ for different sizes $L$ as a function of the occupation probability $p$ for site percolation on the triangular lattice. Inset: The dependence of $p_{eb}(L)$ on $L^{-1}$ for site percolation on the triangular lattice. $p_{eb}(L)$ (circles) is obtained from the crossing point of $U_{L}$ and $U_{2L}$ for each pair of successive $L$ values. Here we consider $L = 64,128,256,512,1024,2048,4096,\mbox{ and }8192$. Extrapolating through the data points (blue line) to the thermodynamic limit, we obtain $p_{eb}(\infty) = 0.7065\pm0.0004$ (red dashed line).}
\label{fig03}
\end{figure}

We define the quantity $M_{eb}$ as the mass of the elastic backbone, i.e., the number of sites that belong to the elastic backbone, and its density as $m_{L}(p) = \left\langle M_{eb}\right\rangle/N$, where $N$ is the total number of sites of the lattice. In Fig.~\ref{fig02} we plot $m_{L}(p)$ against $p > p_c$ for site percolation on the triangular lattice and find that there exists a clear phase transition at a value $p_{eb} > p_c$ at which the elastic backbone becomes dense, with $m$ acting as order parameter of this transition. Using finite size scaling, $p_{eb}$ can be determined more precisely as shown in the inset, yielding also an estimate for the exponent $\beta_{eb}/\nu_{eb} = 0.25\pm 0.02$. The threshold $p_{eb}$ can be determined even more precisely using Binder's cumulant defined as
\begin{equation}
U_{L}(p) = 1 - \frac{\left\langle m_{eb}^{4} \right\rangle}{3 \left\langle m_{eb}^{2} \right\rangle^{2}}.
\end{equation}
In Fig.~\ref{fig03} we show the analysis of Binder's cumulant for site percolation on the triangular lattice, where we obtain $p_{eb} = 0.7065\pm 0.0004$. Moreover, by applying the same analysis, we find for site percolation $p_{eb} = 1$ on the normal square lattice and $p_{eb} = 0.7055 \pm 0.0005$ on the tilted square lattice. For bond percolation, we obtain $p_{eb} = 0.8030\pm 0.0005$ on the tilted square lattice and $p_{eb} = 0.6450\pm0.0004$ on the triangular lattice. At $p=1$, for the square lattice, starting from a given node in the border, there is just one shortest path of size $L$ reaching the other side. In this way, the square lattice leaves the dense phase for any fraction of nodes removed. For the triangular and tilted lattices, on the other hand, there are $2^L$ different shortest paths (of size $L$) reaching the other side. This much larger number of shortest paths increases the mass of the elastic backbone, and therefore the triangular and tilted lattices remain dense, even after some sites have been removed. It should be noted that  a similar dependence on the lattice is observed in other models. In the case of Rigidity Percolation~\cite{kantorPRL1984,bresserPRL1986,moukarzelPRL1997,thorpePRL2015}, a transition on the triangular lattice is observed at finite $p$, while rigidity is only attained at $p=1$ for the square lattice~\cite{moukarzelPRL1998,souslovPRL2009,maoPRL2010}.

\begin{figure}[t]
\includegraphics*[width=\columnwidth]{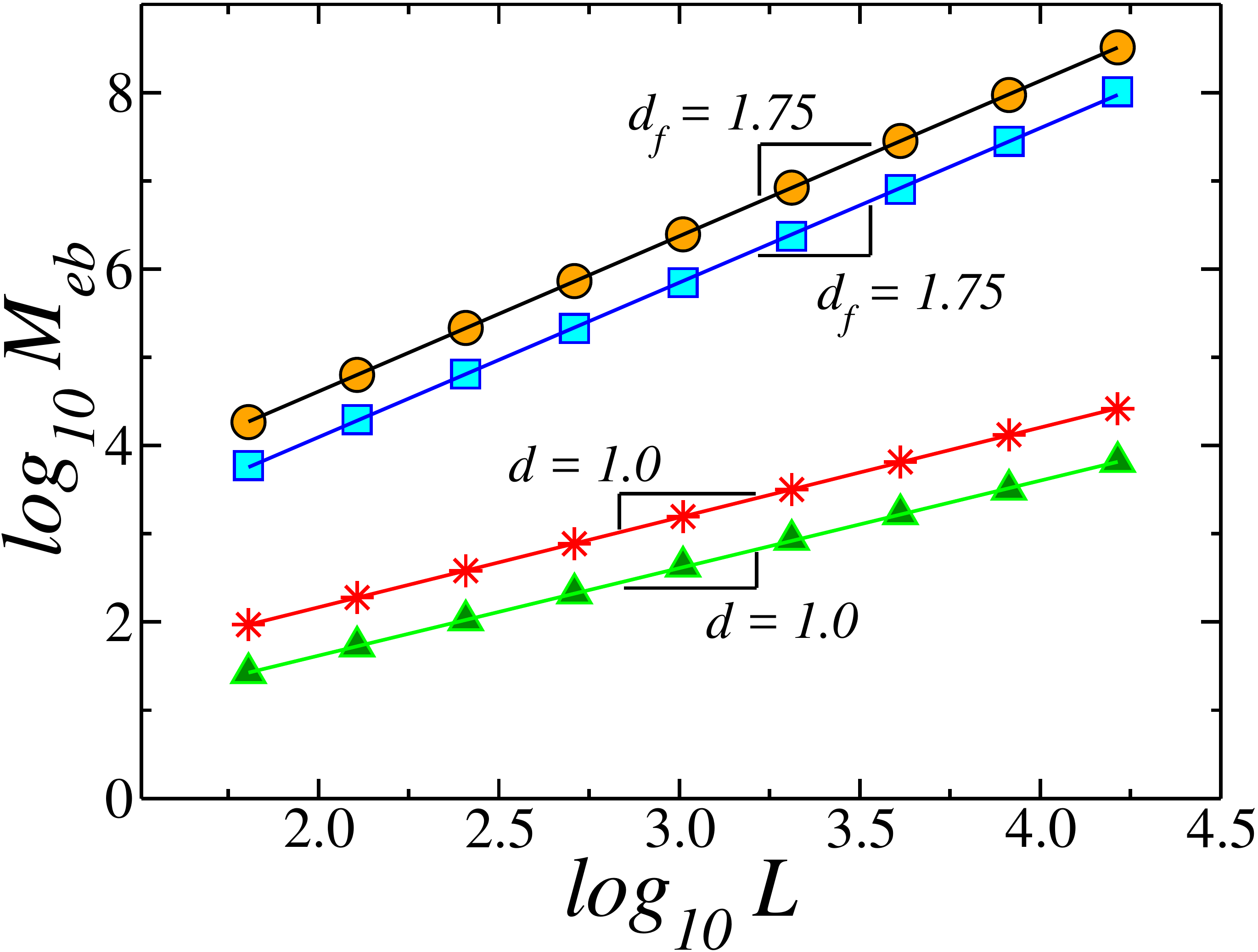}
\caption{Logarithmic plot of the mass $M_{eb}$ of the elastic backbone as a function of the lattice size $L$ for site percolation at $p_{eb}$ on the tilted square lattice (orange circles) and on the triangular lattice (blue squares). The same is shown for the value of $p=0.63$, between $p_{c}$ and $p_{eb}$, on the tilted square lattice for site percolation (red stars) and for bond percolation (green triangles), with $L$ ranging from $64$ to $16384$ sites. At $p=p_{eb}$, the least-squares fit to the data of a power law, $M_{eb}\sim L^{d_{f}}$, gives the exponent $d_{f} = 1.7500\pm0.0003$ for the tilted square lattice (black line) and $d_{f} = 1.7500\pm 0.0002$ for the triangular lattice (blue line). At $p = 0.63$ on the tilted square lattice, the least-squares fit to the data of a power law, $M_{eb}\sim L^{d}$, gives the exponent $d = 1.0000\pm 0.0002$ for site percolation (red line) and  $d = 1.0000\pm 0.0001$ for bond percolation (green line). In all cases, the errors are smaller than the symbols.}
\label{fig04}
\end{figure}

\begin{figure}[t]
\includegraphics*[width=\columnwidth]{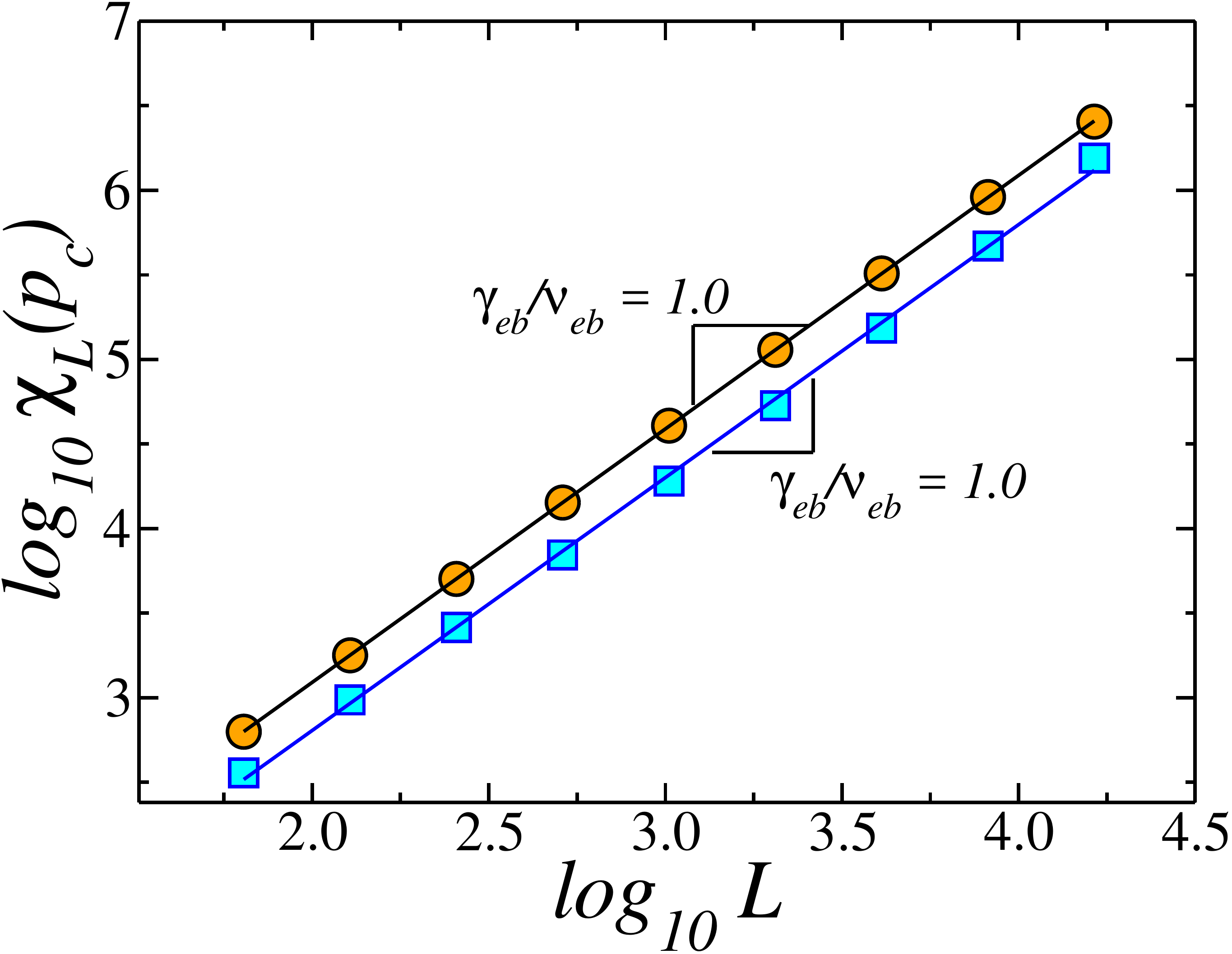}
\caption{Logarithmic plot of the susceptibility $\chi_L$ of the elastic backbone as a function of the lattice size $L$ for site percolation at $p_{eb}$, with $L$ ranging from $64$ to $16384$ sites. The least-squares fit to the data of a power law, $\chi(p_{eb})\sim L^{\gamma/\nu}$, gives the exponent $\gamma/\nu = 1.00\pm 0.01$ for the tilted square lattice (orange circles) and $\gamma/\nu = 1.00\pm 0.02$ for the triangular lattice (blue squares).}
\label{fig05}
\end{figure}

\begin{figure}[t]
\includegraphics*[width=\columnwidth]{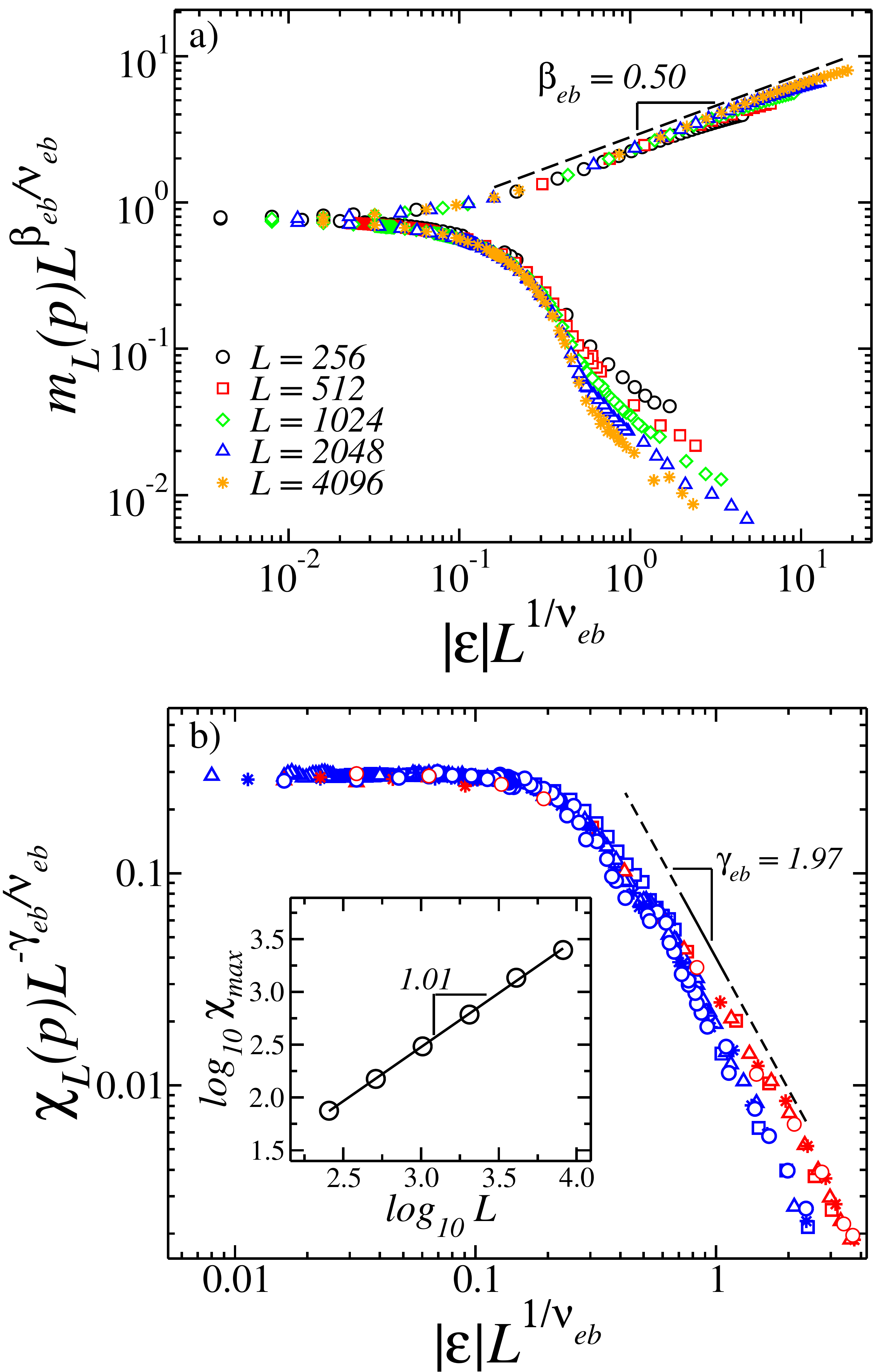}
\caption{(a) Results from the finite-size scaling analysis of the order parameter $m_L$ obtained for site percolation on the triangular lattice. The growing and decaying curves correspond to values above and below $p_{eb}$, respectively. The dashed line is the least-squares fit to data in the scaling region above $p_{eb}$ for Eq.~\ref{eq01}. (b) The same as in (a), but for the susceptibility $\chi_L$. Blue symbols for $p<p_{eb}$ and red symbols for $p > p_{eb}$. Here, the dashed line has slope $1.97$. Inset of $(b)$: Log-log plot of the maximum of $\chi_{L}$ as a function of system size.}
\label{fig06}
\end{figure}

The fractal dimension $d_{f}$ of the elastic backbone is obtained directly from the double-logarithmic plot of the mass of the elastic backbone $M_{eb}$ against the linear size $L$ of the
considered lattice, at the threshold $p_{eb}$. We find that the elastic backbone is fractal with a fractal dimension $d_f = 1.750\pm 0.003$ for all studied models, while for $p_c < p < p_{eb}$ it is one-dimensional, as shown in Fig.~\ref{fig04}. In this case, although the spanning cluster is dense, the existence of various holes prevents the effective coalescence of the shortest paths present in the system, therefore leading to an elastic backbone which is a fractal.  At the threshold $p_{eb}$, the exponent $\beta/\nu$ is then obtained from the relation $\beta/\nu = d - d_{f}$, resulting, for all studied models, in $\beta/\nu = 0.250\pm 0.003$. Moreover, the response function of the order parameter, i.e. what would correspond in magnetic systems to  the susceptibility, $\chi = N(\left\langle m_{eb}^2\right\rangle - \left\langle m_{eb} \right\rangle^{2})$, diverges at the critical threshold $p_{eb}$ with an exponent $\gamma_{eb}/\nu_{eb} = 1.00\pm0.02$ for all models considered, as shown in Fig.~\ref{fig05} for site percolation on the tilted square and triangular lattices. Finally, we also perform a full finite-size scaling analysis for $m_L(p)$ and for $\chi_L(p)$ of the form, 
\begin{equation}
m_{L}(p) = L^{-\beta_{eb}/\nu_{eb}}\widetilde{m}(\varepsilon L^{1/\nu_{eb}}),
\label{eq01} 
\end{equation}

\begin{equation}
\chi_{L}(p) = L^{\gamma_{eb}/\nu_{eb}}\widetilde{\chi}(\varepsilon L^{1/\nu_{eb}}),
\label{eq02} 
\end{equation}
where $\varepsilon = (p - p_{eb})$ is the distance from the critical threshold. The exponents $\beta_{eb}/\nu_{eb}$, $\gamma_{eb}/\nu_{eb}$, and $\nu_{eb}$ are, respectively, associated with the decay of the order parameter, the divergence of the susceptibility, and the finite-size effects. 

As shown in Fig.~\ref{fig06}, for the specific case of site percolation on the triangular lattice, we obtain excellent data collapse for values $\beta_{eb} = 0.50\pm 0.03$, $\gamma_{eb} = 1.97\pm 0.05$  and $\nu_{eb} = 2.00\pm 0.04$. We note that hyperscaling relation $2\beta_{eb} + \gamma_{eb} = d\nu_{eb}$ is violated~\cite{binderJStat1989,binderPRE2010}. Similar data collapse with the same exponents have been found for the other considered lattices.

\begin{figure}[h]
\includegraphics*[width=\columnwidth]{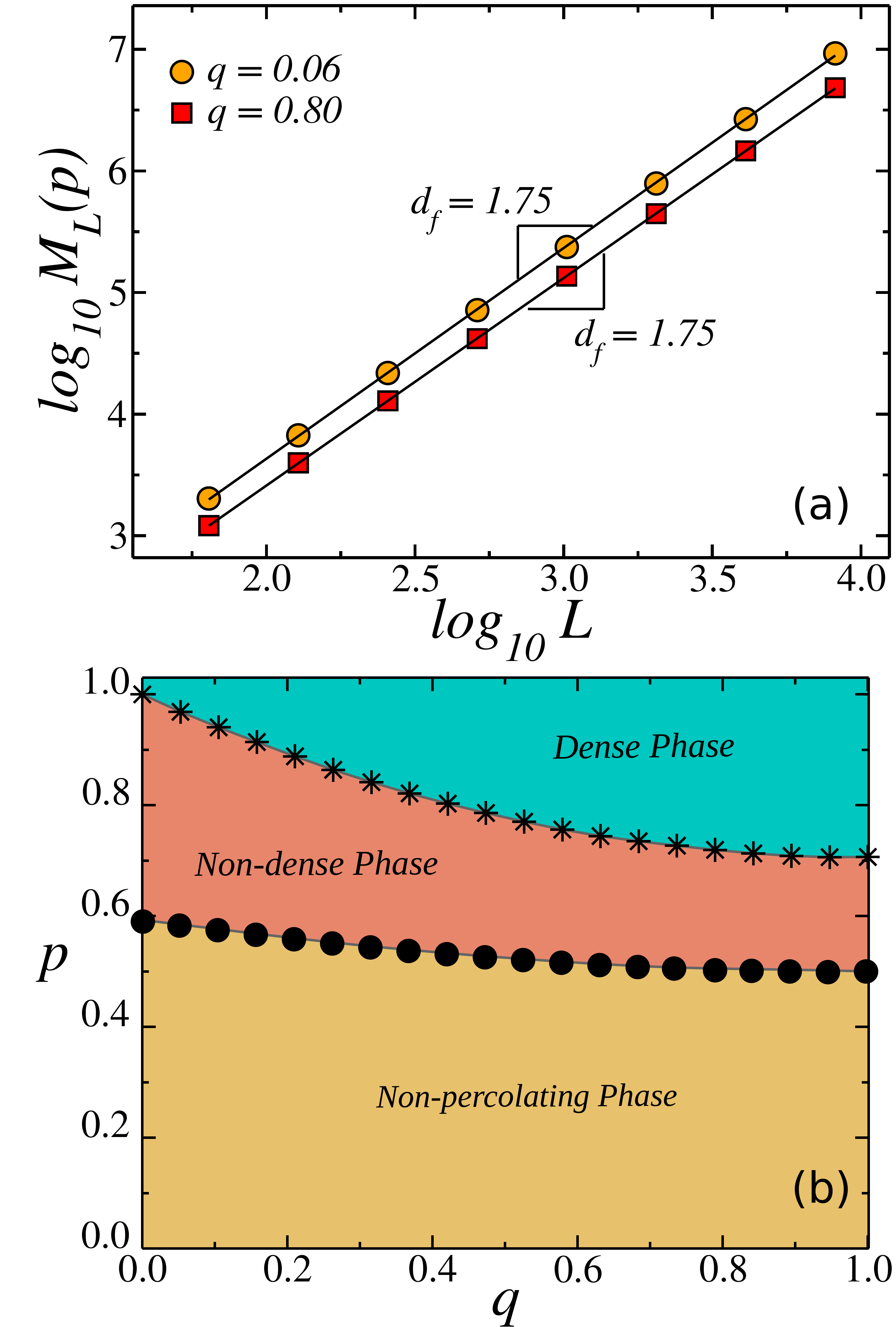}
\caption{ (a) Logarithmic plot of the mass $M_{eb}$ of the elastic backbone as a function of the lattice size $L$ for site percolation at $p_{eb}(q)$ for two different values of $q$. The least-squares fit to the data of a power law, $M_{eb}(p_{eb}(q))\sim L^{d_{f}}$, gives the exponent $d_{f} = 1.750\pm 0.005$ for $q=0.06$ (orange circles) and $d_{f} = 1.750\pm0.003$ for $q = 0.80$ (red squares). The results show that the fractal dimension is within error bars the same for both cases. (b) Phase diagram between percolation probability $p$ and the density of diagonals $q$. Three phases can be identified. The non-percolating phase, where the spanning cluster is absent, bounded on top by the curve $p_{c}(q)$ (filled circles) which represents the classical percolation thresholds. The non-dense phase is bounded by the curve $p_{c}(q)$ on the bottom and the curve $p_{eb}(q)$ on the top (stars) that is the critical line along which the order parameter $m_{L}(p)$ vanishes and the elastic backbone is fractal. Finally, in the dense phase on the top the elastic backbone becomes dense.}
\label{fig07}
\end{figure}

We also studied the elastic backbone transitions on a lattice model that mixes the features of normal square and triangular lattices, by adding to the normal square lattice with probability $q$ some additional diagonals going from top-left to bottom-right. In this way, for $q = 0$, we obtain a normal square lattice and, for $q = 1$, a triangular lattice. For every value of $q > 0$, we found a value of $p_c (q) < 1$ at which the elastic backbone becomes dense. In Fig.~\ref{fig07}a we show the finite-size scaling of the mass $M_{eb}$ of the elastic backbone for site percolation at $p_{eb}(q)$, for $q = 0.06$ and $q = 0.80$. The results show that the fractal dimension is within error bars the same for both cases. Moreover, we calculated these fractal dimensions at $p_{c}(q)$ for different values of $q>0$ and found that they do not depend on $q$. This underlines, on one hand, the universality of $d_{eb}$ and questions, on the other hand, any relation to Rigidity Percolation~\cite{kantorPRL1984,bresserPRL1986,moukarzelPRL1997,thorpePRL2015}, which does exhibit a transition at an intermediate value of $q$. In Fig.~\ref{fig07}b, we show the phase diagram between percolation probability $p$ and the density of diagonals $q$, where three phases are identified. In the non-percolating phase, the spanning cluster is absent and consequently the elastic backbone does not percolate. The non-percolating phase is bounded by the curve $p_{c}(q)$ which represents the classical percolation thresholds. The non-dense phase is characterized by a linear scaling with $L$ of the elastic backbone and it is bounded by the curves $p_{c}(q)$ on the bottom and $p_{eb}(q)$ on the top, which defines the critical line along which the order parameter $m_{L}(p)$ vanishes and the elastic backbone is fractal. In the dense phase, the dimension of the elastic backbone is equal to the dimension of the lattice considered.

Concluding, we discovered that the mass of the elastic backbone serves as order parameter for a new transition within the connected phase of classical percolation, exhibiting a new set of critical exponents  $\beta_{eb} \approx 1/2$, $\gamma_{eb} \approx  2$, $\nu_{eb} \approx 2$, and $d_f \approx 7/4$ in two dimensions. Interestingly, however, hyperscaling is violated, being this to our knowledge the first example for a violation of hyperscaling in a purely geometrical model. It would be also interesting to investigate higher dimensions and try to formulate a mean-field approximation. Similar transitions for elastic backbones could be expected  in models with tunable disorder~\cite{havlinPRL1997,havlinPRL2003,andradePRL2009,pradhan2010failure,andradePRL2011,moreira2012fracturing,zapperiPRL2013,oliveiraPRL2014,sampaioPRL2016}. 

Our findings have direct consequences to the stretching of random fibrous materials like biological tissues: when the first restoring force is felt, the resistance will grow very gently with displacement below the threshold $p_{eb}$, while above $p_{eb}$ the system will then to be instead very stiff. It is thus a transition between two very different stress-strain relations for a damaged tissue. 

Furthermore, there exists an interesting similarity between the elastic backbone percolation, as introduced here, and the phase transition associated to Rigidity Percolation, since the stressed backbone~\cite{moukarzelPRL1995,thorpePRL1998,moukarzelPRE1999} is a fractal whose dimension ($d_f = 1.78\pm 0.02$) is very close to the fractal dimension of the elastic backbone at $p_{eb}$. For non-tilted square lattices, in both cases, the transition is shifted to $p = 1$~\cite{moukarzelPRL1998,souslovPRL2009,maoPRL2010}. Nevertheless the two transitions describe different phenomena, since they are in general located at different thresholds values and have a different set of critical exponents.      

\begin{acknowledgments}
We thank the Brazilian agencies CNPq, CAPES, FUNCAP, the National Institute of Science and Technology for Complex Systems, and the European Research Council (ERC) Advanced Grant 319968 FlowCCS for financial support.
\end{acknowledgments}


\begin{thebibliography}{39}
\expandafter\ifx\csname natexlab\endcsname\relax\def\natexlab#1{#1}\fi
\expandafter\ifx\csname bibnamefont\endcsname\relax
  \def\bibnamefont#1{#1}\fi
\expandafter\ifx\csname bibfnamefont\endcsname\relax
  \def\bibfnamefont#1{#1}\fi
\expandafter\ifx\csname citenamefont\endcsname\relax
  \def\citenamefont#1{#1}\fi
\expandafter\ifx\csname url\endcsname\relax
  \def\url#1{\texttt{#1}}\fi
\expandafter\ifx\csname urlprefix\endcsname\relax\def\urlprefix{URL }\fi
\providecommand{\bibinfo}[2]{#2}
\providecommand{\eprint}[2][]{\url{#2}}

\bibitem[{\citenamefont{Broadbent and
  Hammersley}(1957)}]{broadbent1957percolation}
\bibinfo{author}{\bibfnamefont{S.~R.} \bibnamefont{Broadbent}}
  \bibnamefont{and} \bibinfo{author}{\bibfnamefont{J.~M.}
  \bibnamefont{Hammersley}}, in \emph{\bibinfo{booktitle}{Math. Proc.
  Cambridge}} (\bibinfo{organization}{Cambridge University Press},
  \bibinfo{year}{1957}), vol.~\bibinfo{volume}{53}, pp.
  \bibinfo{pages}{629--641}.

\bibitem[{\citenamefont{Kirkpatrick}(1973)}]{kirkpatrick1973percolation}
\bibinfo{author}{\bibfnamefont{S.}~\bibnamefont{Kirkpatrick}},
  \bibinfo{journal}{Rev. Mod. Phys.} \textbf{\bibinfo{volume}{45}},
  \bibinfo{pages}{574} (\bibinfo{year}{1973}).

\bibitem[{\citenamefont{Stauffer and Aharony}(1985)}]{stauferbook1985}
\bibinfo{author}{\bibfnamefont{D.}~\bibnamefont{Stauffer}} \bibnamefont{and}
  \bibinfo{author}{\bibfnamefont{A.}~\bibnamefont{Aharony}},
  \bibinfo{journal}{Tailor \& Francis, London}  (\bibinfo{year}{1985}).

\bibitem[{\citenamefont{Sahimi}(1994)}]{sahimiBook1994}
\bibinfo{author}{\bibfnamefont{M.}~\bibnamefont{Sahimi}},
  \emph{\bibinfo{title}{Applications of Percolation Theory}}
  (\bibinfo{publisher}{Taylor \& Francis Group}, \bibinfo{address}{London, UK},
  \bibinfo{year}{1994}).

\bibitem[{\citenamefont{Herrmann and Stanley}(1988)}]{herrmann1988fractal}
\bibinfo{author}{\bibfnamefont{H.~J.} \bibnamefont{Herrmann}} \bibnamefont{and}
  \bibinfo{author}{\bibfnamefont{H.~E.} \bibnamefont{Stanley}},
  \bibinfo{journal}{J. Phys. A-Math. Gen.} \textbf{\bibinfo{volume}{21}},
  \bibinfo{pages}{L829} (\bibinfo{year}{1988}).

\bibitem[{\citenamefont{Grassberger}(1992)}]{grassberger1992}
\bibinfo{author}{\bibfnamefont{P.}~\bibnamefont{Grassberger}},
  \bibinfo{journal}{J. Phys. A-Math. Gen.} \textbf{\bibinfo{volume}{25}},
  \bibinfo{pages}{5475} (\bibinfo{year}{1992}).

\bibitem[{\citenamefont{Dokholyan et~al.}(1998)\citenamefont{Dokholyan, Lee,
  Buldyrev, Havlin, King, and Stanley}}]{havlinJStat1998}
\bibinfo{author}{\bibfnamefont{N.~V.} \bibnamefont{Dokholyan}},
  \bibinfo{author}{\bibfnamefont{Y.}~\bibnamefont{Lee}},
  \bibinfo{author}{\bibfnamefont{S.~V.} \bibnamefont{Buldyrev}},
  \bibinfo{author}{\bibfnamefont{S.}~\bibnamefont{Havlin}},
  \bibinfo{author}{\bibfnamefont{P.~R.} \bibnamefont{King}}, \bibnamefont{and}
  \bibinfo{author}{\bibfnamefont{H.~E.} \bibnamefont{Stanley}},
  \bibinfo{journal}{J. Stat. Phys.} \textbf{\bibinfo{volume}{93}},
  \bibinfo{pages}{603} (\bibinfo{year}{1998}).

\bibitem[{\citenamefont{Newman and Ziff}(2000)}]{ziffPRL2000}
\bibinfo{author}{\bibfnamefont{M.~E.~J.} \bibnamefont{Newman}}
  \bibnamefont{and} \bibinfo{author}{\bibfnamefont{R.~M.} \bibnamefont{Ziff}},
  \bibinfo{journal}{Phys. Rev. Lett.} \textbf{\bibinfo{volume}{85}},
  \bibinfo{pages}{4104} (\bibinfo{year}{2000}).

\bibitem[{\citenamefont{Herrmann et~al.}(1984)\citenamefont{Herrmann, Hong, and
  Stanley}}]{herrmann1984backbone}
\bibinfo{author}{\bibfnamefont{H.~J.} \bibnamefont{Herrmann}},
  \bibinfo{author}{\bibfnamefont{D.~C.} \bibnamefont{Hong}}, \bibnamefont{and}
  \bibinfo{author}{\bibfnamefont{H.~E.} \bibnamefont{Stanley}},
  \bibinfo{journal}{J. Phys. A-Math. Gen.} \textbf{\bibinfo{volume}{17}},
  \bibinfo{pages}{L261} (\bibinfo{year}{1984}).

\bibitem[{\citenamefont{Jacobs and Thorpe}(1995)}]{thorpePRL1995}
\bibinfo{author}{\bibfnamefont{D.~J.} \bibnamefont{Jacobs}} \bibnamefont{and}
  \bibinfo{author}{\bibfnamefont{M.~F.} \bibnamefont{Thorpe}},
  \bibinfo{journal}{Phys. Rev. Lett.} \textbf{\bibinfo{volume}{75}},
  \bibinfo{pages}{4051} (\bibinfo{year}{1995}).

\bibitem[{\citenamefont{de~Souza and Harrowell}(2009)}]{souzaPNAS2009}
\bibinfo{author}{\bibfnamefont{V.~K.} \bibnamefont{de~Souza}} \bibnamefont{and}
  \bibinfo{author}{\bibfnamefont{P.}~\bibnamefont{Harrowell}},
  \bibinfo{journal}{Proc. Natl. Acad. Sci. USA} \textbf{\bibinfo{volume}{106}},
  \bibinfo{pages}{15136} (\bibinfo{year}{2009}).

\bibitem[{\citenamefont{Zhang et~al.}(2017)\citenamefont{Zhang, Rocklin,
  Sander, and Mao}}]{PhysRevMaterials}
\bibinfo{author}{\bibfnamefont{L.}~\bibnamefont{Zhang}},
  \bibinfo{author}{\bibfnamefont{D.~Z.} \bibnamefont{Rocklin}},
  \bibinfo{author}{\bibfnamefont{L.~M.} \bibnamefont{Sander}},
  \bibnamefont{and} \bibinfo{author}{\bibfnamefont{X.}~\bibnamefont{Mao}},
  \bibinfo{journal}{Phys. Rev. Materials} \textbf{\bibinfo{volume}{1}},
  \bibinfo{pages}{052602} (\bibinfo{year}{2017}).

\bibitem[{\citenamefont{Bates et~al.}(1994)\citenamefont{Bates, Maksym,
  Navajas, and Suki}}]{bates1994lung}
\bibinfo{author}{\bibfnamefont{J.}~\bibnamefont{Bates}},
  \bibinfo{author}{\bibfnamefont{G.}~\bibnamefont{Maksym}},
  \bibinfo{author}{\bibfnamefont{D.}~\bibnamefont{Navajas}}, \bibnamefont{and}
  \bibinfo{author}{\bibfnamefont{B.}~\bibnamefont{Suki}},
  \bibinfo{journal}{Ann. Biomed. Eng.} \textbf{\bibinfo{volume}{22}},
  \bibinfo{pages}{674} (\bibinfo{year}{1994}).

\bibitem[{\citenamefont{Yuan et~al.}(1997)\citenamefont{Yuan, Ingenito, and
  Suki}}]{yuan1997dynamic}
\bibinfo{author}{\bibfnamefont{H.}~\bibnamefont{Yuan}},
  \bibinfo{author}{\bibfnamefont{E.~P.} \bibnamefont{Ingenito}},
  \bibnamefont{and} \bibinfo{author}{\bibfnamefont{B.}~\bibnamefont{Suki}},
  \bibinfo{journal}{J. Appl. Physiol.} \textbf{\bibinfo{volume}{83}},
  \bibinfo{pages}{1420} (\bibinfo{year}{1997}).

\bibitem[{\citenamefont{Baish and Jain}(2000)}]{baish2000fractals}
\bibinfo{author}{\bibfnamefont{J.~W.} \bibnamefont{Baish}} \bibnamefont{and}
  \bibinfo{author}{\bibfnamefont{R.~K.} \bibnamefont{Jain}},
  \bibinfo{journal}{Cancer Res.} \textbf{\bibinfo{volume}{60}},
  \bibinfo{pages}{3683} (\bibinfo{year}{2000}).

\bibitem[{\citenamefont{Suki}(2002)}]{suki2002fluctuations}
\bibinfo{author}{\bibfnamefont{B.}~\bibnamefont{Suki}}, \bibinfo{journal}{Am.
  J. Respir. Crit. Care Med.} \textbf{\bibinfo{volume}{166}},
  \bibinfo{pages}{133} (\bibinfo{year}{2002}).

\bibitem[{\citenamefont{Ritter et~al.}(2009)\citenamefont{Ritter, Jesudason,
  Majumdar, Stamenovi{\'c}, Buczek-Thomas, Stone, Nugent, and
  Suki}}]{ritter2009zipper}
\bibinfo{author}{\bibfnamefont{M.~C.} \bibnamefont{Ritter}},
  \bibinfo{author}{\bibfnamefont{R.}~\bibnamefont{Jesudason}},
  \bibinfo{author}{\bibfnamefont{A.}~\bibnamefont{Majumdar}},
  \bibinfo{author}{\bibfnamefont{D.}~\bibnamefont{Stamenovi{\'c}}},
  \bibinfo{author}{\bibfnamefont{J.~A.} \bibnamefont{Buczek-Thomas}},
  \bibinfo{author}{\bibfnamefont{P.~J.} \bibnamefont{Stone}},
  \bibinfo{author}{\bibfnamefont{M.~A.} \bibnamefont{Nugent}},
  \bibnamefont{and} \bibinfo{author}{\bibfnamefont{B.}~\bibnamefont{Suki}},
  \bibinfo{journal}{Proc. Natl. Acad. Sci. USA} \textbf{\bibinfo{volume}{106}},
  \bibinfo{pages}{1081} (\bibinfo{year}{2009}).

\bibitem[{\citenamefont{Lennon et~al.}(2015)\citenamefont{Lennon, Cianci,
  Cipriani, Hensing, Zhang, Chen, Murgu, Vokes, Vannier, and
  Salgia}}]{lennon2015lung}
\bibinfo{author}{\bibfnamefont{F.~E.} \bibnamefont{Lennon}},
  \bibinfo{author}{\bibfnamefont{G.~C.} \bibnamefont{Cianci}},
  \bibinfo{author}{\bibfnamefont{N.~A.} \bibnamefont{Cipriani}},
  \bibinfo{author}{\bibfnamefont{T.~A.} \bibnamefont{Hensing}},
  \bibinfo{author}{\bibfnamefont{H.~J.} \bibnamefont{Zhang}},
  \bibinfo{author}{\bibfnamefont{C.-T.} \bibnamefont{Chen}},
  \bibinfo{author}{\bibfnamefont{S.~D.} \bibnamefont{Murgu}},
  \bibinfo{author}{\bibfnamefont{E.~E.} \bibnamefont{Vokes}},
  \bibinfo{author}{\bibfnamefont{M.~W.} \bibnamefont{Vannier}},
  \bibnamefont{and} \bibinfo{author}{\bibfnamefont{R.}~\bibnamefont{Salgia}},
  \bibinfo{journal}{Nat. Rev. Clin. Oncol.} \textbf{\bibinfo{volume}{12}},
  \bibinfo{pages}{664} (\bibinfo{year}{2015}).

\bibitem[{\citenamefont{Kantor and Webman}(1984)}]{kantorPRL1984}
\bibinfo{author}{\bibfnamefont{Y.}~\bibnamefont{Kantor}} \bibnamefont{and}
  \bibinfo{author}{\bibfnamefont{I.}~\bibnamefont{Webman}},
  \bibinfo{journal}{Phys. Rev. Lett.} \textbf{\bibinfo{volume}{52}},
  \bibinfo{pages}{1891} (\bibinfo{year}{1984}).

\bibitem[{\citenamefont{Bresser et~al.}(1986)\citenamefont{Bresser, Boolchand,
  and Suranyi}}]{bresserPRL1986}
\bibinfo{author}{\bibfnamefont{W.}~\bibnamefont{Bresser}},
  \bibinfo{author}{\bibfnamefont{P.}~\bibnamefont{Boolchand}},
  \bibnamefont{and} \bibinfo{author}{\bibfnamefont{P.}~\bibnamefont{Suranyi}},
  \bibinfo{journal}{Phys. Rev. Lett.} \textbf{\bibinfo{volume}{56}},
  \bibinfo{pages}{2493} (\bibinfo{year}{1986}).

\bibitem[{\citenamefont{Moukarzel et~al.}(1997)\citenamefont{Moukarzel,
  Duxbury, and Leath}}]{moukarzelPRL1997}
\bibinfo{author}{\bibfnamefont{C.}~\bibnamefont{Moukarzel}},
  \bibinfo{author}{\bibfnamefont{P.~M.} \bibnamefont{Duxbury}},
  \bibnamefont{and} \bibinfo{author}{\bibfnamefont{P.}~\bibnamefont{Leath}},
  \bibinfo{journal}{Phys. Rev. Lett.} \textbf{\bibinfo{volume}{78}},
  \bibinfo{pages}{1480} (\bibinfo{year}{1997}).

\bibitem[{\citenamefont{Ellenbroek et~al.}(2015)\citenamefont{Ellenbroek, Hagh,
  Kumar, Thorpe, and Van~Hecke}}]{thorpePRL2015}
\bibinfo{author}{\bibfnamefont{W.~G.} \bibnamefont{Ellenbroek}},
  \bibinfo{author}{\bibfnamefont{V.~F.} \bibnamefont{Hagh}},
  \bibinfo{author}{\bibfnamefont{A.}~\bibnamefont{Kumar}},
  \bibinfo{author}{\bibfnamefont{M.~F.} \bibnamefont{Thorpe}},
  \bibnamefont{and}
  \bibinfo{author}{\bibfnamefont{M.}~\bibnamefont{Van~Hecke}},
  \bibinfo{journal}{Phys. Rev. Lett.} \textbf{\bibinfo{volume}{114}},
  \bibinfo{pages}{135501} (\bibinfo{year}{2015}).

\bibitem[{\citenamefont{Moukarzel}(1998)}]{moukarzelPRL1998}
\bibinfo{author}{\bibfnamefont{C.~F.} \bibnamefont{Moukarzel}},
  \bibinfo{journal}{Phys. Rev. Lett.} \textbf{\bibinfo{volume}{81}},
  \bibinfo{pages}{1634} (\bibinfo{year}{1998}).

\bibitem[{\citenamefont{Souslov et~al.}(2009)\citenamefont{Souslov, Liu, and
  Lubensky}}]{souslovPRL2009}
\bibinfo{author}{\bibfnamefont{A.}~\bibnamefont{Souslov}},
  \bibinfo{author}{\bibfnamefont{A.~J.} \bibnamefont{Liu}}, \bibnamefont{and}
  \bibinfo{author}{\bibfnamefont{T.~C.} \bibnamefont{Lubensky}},
  \bibinfo{journal}{Phys. Rev. Lett.} \textbf{\bibinfo{volume}{103}},
  \bibinfo{pages}{205503} (\bibinfo{year}{2009}).

\bibitem[{\citenamefont{Mao et~al.}(2010)\citenamefont{Mao, Xu, and
  Lubensky}}]{maoPRL2010}
\bibinfo{author}{\bibfnamefont{X.}~\bibnamefont{Mao}},
  \bibinfo{author}{\bibfnamefont{N.}~\bibnamefont{Xu}}, \bibnamefont{and}
  \bibinfo{author}{\bibfnamefont{T.~C.} \bibnamefont{Lubensky}},
  \bibinfo{journal}{Phys. Rev. Lett.} \textbf{\bibinfo{volume}{104}},
  \bibinfo{pages}{085504} (\bibinfo{year}{2010}).

\bibitem[{\citenamefont{Binder and Wang}(1989)}]{binderJStat1989}
\bibinfo{author}{\bibfnamefont{K.}~\bibnamefont{Binder}} \bibnamefont{and}
  \bibinfo{author}{\bibfnamefont{J.-S.} \bibnamefont{Wang}},
  \bibinfo{journal}{J. Stat. Phys.} \textbf{\bibinfo{volume}{55}},
  \bibinfo{pages}{87} (\bibinfo{year}{1989}).

\bibitem[{\citenamefont{Vink et~al.}(2010)\citenamefont{Vink, Fischer, and
  Binder}}]{binderPRE2010}
\bibinfo{author}{\bibfnamefont{R.~L.~C.} \bibnamefont{Vink}},
  \bibinfo{author}{\bibfnamefont{T.}~\bibnamefont{Fischer}}, \bibnamefont{and}
  \bibinfo{author}{\bibfnamefont{K.}~\bibnamefont{Binder}},
  \bibinfo{journal}{Phys. Rev. E} \textbf{\bibinfo{volume}{82}},
  \bibinfo{pages}{051134} (\bibinfo{year}{2010}).

\bibitem[{\citenamefont{Porto et~al.}(1997)\citenamefont{Porto, Havlin,
  Schwarzer, and Bunde}}]{havlinPRL1997}
\bibinfo{author}{\bibfnamefont{M.}~\bibnamefont{Porto}},
  \bibinfo{author}{\bibfnamefont{S.}~\bibnamefont{Havlin}},
  \bibinfo{author}{\bibfnamefont{S.}~\bibnamefont{Schwarzer}},
  \bibnamefont{and} \bibinfo{author}{\bibfnamefont{A.}~\bibnamefont{Bunde}},
  \bibinfo{journal}{Phys. Rev. Lett.} \textbf{\bibinfo{volume}{79}},
  \bibinfo{pages}{4060} (\bibinfo{year}{1997}).

\bibitem[{\citenamefont{Braunstein et~al.}(2003)\citenamefont{Braunstein,
  Buldyrev, Cohen, Havlin, and Stanley}}]{havlinPRL2003}
\bibinfo{author}{\bibfnamefont{L.~A.} \bibnamefont{Braunstein}},
  \bibinfo{author}{\bibfnamefont{S.~V.} \bibnamefont{Buldyrev}},
  \bibinfo{author}{\bibfnamefont{R.}~\bibnamefont{Cohen}},
  \bibinfo{author}{\bibfnamefont{S.}~\bibnamefont{Havlin}}, \bibnamefont{and}
  \bibinfo{author}{\bibfnamefont{H.~E.} \bibnamefont{Stanley}},
  \bibinfo{journal}{Phys. Rev. Lett.} \textbf{\bibinfo{volume}{91}},
  \bibinfo{pages}{168701} (\bibinfo{year}{2003}).

\bibitem[{\citenamefont{Andrade et~al.}(2009)\citenamefont{Andrade, Oliveira,
  Moreira, and Herrmann}}]{andradePRL2009}
\bibinfo{author}{\bibfnamefont{J.~S.} \bibnamefont{Andrade}},
  \bibinfo{author}{\bibfnamefont{E.~A.} \bibnamefont{Oliveira}},
  \bibinfo{author}{\bibfnamefont{A.~A.} \bibnamefont{Moreira}},
  \bibnamefont{and} \bibinfo{author}{\bibfnamefont{H.~J.}
  \bibnamefont{Herrmann}}, \bibinfo{journal}{Phys. Rev. Lett.}
  \textbf{\bibinfo{volume}{103}}, \bibinfo{pages}{225503}
  (\bibinfo{year}{2009}).

\bibitem[{\citenamefont{Pradhan et~al.}(2010)\citenamefont{Pradhan, Hansen, and
  Chakrabarti}}]{pradhan2010failure}
\bibinfo{author}{\bibfnamefont{S.}~\bibnamefont{Pradhan}},
  \bibinfo{author}{\bibfnamefont{A.}~\bibnamefont{Hansen}}, \bibnamefont{and}
  \bibinfo{author}{\bibfnamefont{B.~K.} \bibnamefont{Chakrabarti}},
  \bibinfo{journal}{Rev. Mod. Phys.} \textbf{\bibinfo{volume}{82}},
  \bibinfo{pages}{499} (\bibinfo{year}{2010}).

\bibitem[{\citenamefont{Fehr et~al.}(2011)\citenamefont{Fehr, Kadau, Andrade,
  and Herrmann}}]{andradePRL2011}
\bibinfo{author}{\bibfnamefont{E.}~\bibnamefont{Fehr}},
  \bibinfo{author}{\bibfnamefont{D.}~\bibnamefont{Kadau}},
  \bibinfo{author}{\bibfnamefont{J.~S.} \bibnamefont{Andrade}},
  \bibnamefont{and} \bibinfo{author}{\bibfnamefont{H.~J.}
  \bibnamefont{Herrmann}}, \bibinfo{journal}{Phys. Rev. Lett.}
  \textbf{\bibinfo{volume}{106}}, \bibinfo{pages}{048501}
  (\bibinfo{year}{2011}).

\bibitem[{\citenamefont{Moreira et~al.}(2012)\citenamefont{Moreira, Oliveira,
  Hansen, Ara{\'u}jo, Herrmann, and Andrade~Jr}}]{moreira2012fracturing}
\bibinfo{author}{\bibfnamefont{A.~A.} \bibnamefont{Moreira}},
  \bibinfo{author}{\bibfnamefont{C.~L.~N.} \bibnamefont{Oliveira}},
  \bibinfo{author}{\bibfnamefont{A.}~\bibnamefont{Hansen}},
  \bibinfo{author}{\bibfnamefont{N.~A.~M.} \bibnamefont{Ara{\'u}jo}},
  \bibinfo{author}{\bibfnamefont{H.~J.} \bibnamefont{Herrmann}},
  \bibnamefont{and} \bibinfo{author}{\bibfnamefont{J.~S.}
  \bibnamefont{Andrade~Jr}}, \bibinfo{journal}{Phys. Rev. Lett.}
  \textbf{\bibinfo{volume}{109}}, \bibinfo{pages}{255701}
  (\bibinfo{year}{2012}).

\bibitem[{\citenamefont{Shekhawat et~al.}(2013)\citenamefont{Shekhawat,
  Zapperi, and Sethna}}]{zapperiPRL2013}
\bibinfo{author}{\bibfnamefont{A.}~\bibnamefont{Shekhawat}},
  \bibinfo{author}{\bibfnamefont{S.}~\bibnamefont{Zapperi}}, \bibnamefont{and}
  \bibinfo{author}{\bibfnamefont{J.~P.} \bibnamefont{Sethna}},
  \bibinfo{journal}{Phys. Rev. Lett.} \textbf{\bibinfo{volume}{110}},
  \bibinfo{pages}{185505} (\bibinfo{year}{2013}).

\bibitem[{\citenamefont{Oliveira et~al.}(2014)\citenamefont{Oliveira, Morais,
  Moreira, and Andrade}}]{oliveiraPRL2014}
\bibinfo{author}{\bibfnamefont{C.~L.~N.} \bibnamefont{Oliveira}},
  \bibinfo{author}{\bibfnamefont{P.~A.} \bibnamefont{Morais}},
  \bibinfo{author}{\bibfnamefont{A.~A.} \bibnamefont{Moreira}},
  \bibnamefont{and} \bibinfo{author}{\bibfnamefont{J.~S.}
  \bibnamefont{Andrade}}, \bibinfo{journal}{Phys. Rev. Lett.}
  \textbf{\bibinfo{volume}{112}}, \bibinfo{pages}{148701}
  (\bibinfo{year}{2014}).

\bibitem[{\citenamefont{Sampaio~Filho et~al.}(2016)\citenamefont{Sampaio~Filho,
  Moreira, Ara\'ujo, Andrade~Jr, and Herrmann}}]{sampaioPRL2016}
\bibinfo{author}{\bibfnamefont{C.~I.~N.} \bibnamefont{Sampaio~Filho}},
  \bibinfo{author}{\bibfnamefont{A.~A.} \bibnamefont{Moreira}},
  \bibinfo{author}{\bibfnamefont{N.~A.~M.} \bibnamefont{Ara\'ujo}},
  \bibinfo{author}{\bibfnamefont{J.~S.} \bibnamefont{Andrade~Jr}},
  \bibnamefont{and} \bibinfo{author}{\bibfnamefont{H.~J.}
  \bibnamefont{Herrmann}}, \bibinfo{journal}{Phys. Rev. Lett.}
  \textbf{\bibinfo{volume}{117}}, \bibinfo{pages}{275702}
  (\bibinfo{year}{2016}).

\bibitem[{\citenamefont{Moukarzel and Duxbury}(1995)}]{moukarzelPRL1995}
\bibinfo{author}{\bibfnamefont{C.}~\bibnamefont{Moukarzel}} \bibnamefont{and}
  \bibinfo{author}{\bibfnamefont{P.~M.} \bibnamefont{Duxbury}},
  \bibinfo{journal}{Phys. Rev. Lett.} \textbf{\bibinfo{volume}{75}},
  \bibinfo{pages}{4055} (\bibinfo{year}{1995}).

\bibitem[{\citenamefont{Jacobs and Thorpe}(1998)}]{thorpePRL1998}
\bibinfo{author}{\bibfnamefont{D.~J.} \bibnamefont{Jacobs}} \bibnamefont{and}
  \bibinfo{author}{\bibfnamefont{M.~F.} \bibnamefont{Thorpe}},
  \bibinfo{journal}{Phys. Rev. Lett.} \textbf{\bibinfo{volume}{80}},
  \bibinfo{pages}{5451} (\bibinfo{year}{1998}).

\bibitem[{\citenamefont{Moukarzel and Duxbury}(1999)}]{moukarzelPRE1999}
\bibinfo{author}{\bibfnamefont{C.}~\bibnamefont{Moukarzel}} \bibnamefont{and}
  \bibinfo{author}{\bibfnamefont{P.~M.} \bibnamefont{Duxbury}},
  \bibinfo{journal}{Phys. Rev. E} \textbf{\bibinfo{volume}{59}},
  \bibinfo{pages}{2614} (\bibinfo{year}{1999}).

\end{thebibliography}
\end{document}